\documentclass[11pt]{article}
\usepackage{jheppub}
\usepackage{epsfig}
\usepackage{bmpsize}
\usepackage{hyperref}
\usepackage{fancyhdr}
\DeclareGraphicsExtensions{.bmp,.png,.pdf,.jpg,.jpeg}
\usepackage{caption}
\usepackage{subcaption}
\usepackage{float}
\usepackage{amsmath}
\usepackage{amsfonts}
\usepackage{textcomp}
\usepackage{bbold}
\usepackage{amssymb}
\usepackage{pifont}
\usepackage{multicol}
\usepackage{array}
\usepackage{stmaryrd}
\usepackage{bm}
\usepackage{setspace}
\usepackage{braket}
\usepackage{dsfont}

\title{Magnetic dipole moments for composite dark matter}
\author[a,b]{Alfredo Aranda,}
\author[a,c]{Luis Barajas}
\author[a,b,d]{Jose A.R. Cembranos}

\affiliation[a] {Facultad de Ciencias - CUICBAS, Universidad de Colima, C.P. 28045, Colima, M\'exico}
\affiliation[b] {Dual CP Institute of High Energy Physics, C.P. 28045, Colima, M\'exico}
\affiliation[c] {Department of Physics, University at Buffalo, The State University of
New York, Buffalo, NY 14260-1500, USA}
\affiliation[d] {Departamento de F\'{\i}sica Te\'orica I, Universidad Complutense de Madrid, E-28040 Madrid, Spain}

\emailAdd{fefo@ucol.mx}
\emailAdd{luisedua@buffalo.edu}
\emailAdd{cembra@fis.ucm.es }

\abstract{
We study neutral dark matter candidates with a nonzero magnetic dipole moment.
We assume that they are composite states of new fermions related to the strong phase of
a new gauge interaction. In particular, invoking a dark flavor symmetry, we analyze the composition structure of viable candidates depending on the assignations of hypercharge and the multiplets
associated to the fundamental constituents of the extended sector. We determine the magnetic dipole moments for the neutral composite states in terms of their constituents masses.}


\notoc

\begin{document}
\maketitle
\section{Introduction}

Many different observational evidences prove the existence of Dark Matter (DM).
Galaxy clusters and dynamics, structure formation, big-bang
nucleosynthesis, and the cosmic microwave background show that baryons can only
account for a small part of the total matter density of the Universe. Many extensions of the Standard Model (SM) provide viable DM candidates, however no clear evidence for any particular extension has been found. This fact motivates the analysis
of DM properties from a broader approach.

DM is typically assumed to have negligible direct couplings to photons but there are other interesting possibilities~\cite{other} such as Magnetic DM (MDM), i.e.
DM particles with a nonzero magnetic dipole moment ($\mu_{\rm DM}$). This has been explored in different
scenarios with some interesting results. For example in Ref.~\cite{Sigurdson:2004zp} one finds a general study of its
phenomenological signatures and constraints. Ref.~\cite{Masso:2009mu} studies direct detection in experimental observations to constraint different types of MDM. In~\cite{Barger:2010gv} the authors analyze the situation particularly for the CoGeNT data for a DM mass of
several GeVs. Ref.~\cite{Banks:2010eh} remarks similar results for the DAMA signature.
The works in~\cite{Fortin:2011hv} and \cite{DelNobile:2012tx} also analyze direct detection results for MDM, but they compare their conclusions with other constraints from indirect searches and colliders. The indirect detection of MDM is studied in Ref.~\cite{Cline:2012bz} as a possible explanation of the $130$ GeV line observed by Fermi-LAT. The constraining power of
supernova SN 1987A data in order to restrict the viability of light MDM is shown in~\cite{Kadota:2014mea}. A similar analysis for the beam dump experiment E613 is done in Ref.~\cite{Mohanty:2015koa}.

A particle with a permanent ${\mu_{\rm DM}}$ must have a
nonzero spin. In this work we only consider fundamental
spin-1/2 Dirac fermions $\psi_{\rm DM}$, since a
Majorana fermion cannot have this type of moments.
In contrast with the electric dipole moment, the magnetic dipole moment is
an axial vector and can couple to the spin without violating time-reversal
and parity symmetries.  In contrast to charged particles or neutral particles with
electric dipole moments~\cite{Fermi:1947},
particles provided with a $\mu_{\rm DM}$ do not have the ability to form atom-like bound states
with other charged particles or with each other. This fact changes completely the
phenomenology of MDM.

The magnetic dipole moment is expected to be small enough to satisfy perturbative constraints:
${\mu_{\rm DM}} \lesssim e \, m_{\rm DM}^{-1}\simeq 2\, (m_e/m_{\rm DM})$,
where $m_{\rm DM}$ is the mass of the DM particle.
A more rigorous bound can be imposed by
unitarity arguments. Indeed, the total $s$-wave annihilation cross
section must be $\sigma \lesssim 4\pi/m_{\rm DM}^2$ \cite{Griest:1989wd}.
By using the expression for two photons annihilation \cite{Sigurdson:2004zp,Fortin:2011hv},
it is possible to find ${\mu_{\rm DM}}\, m_{\rm DM} \lesssim 20\, (m_e/m_{\rm DM})$.

The viability of  MDM can be divided in three different regions depending on its mass: if $m_{\rm DM} \lesssim 10$~MeV, the constraints on
additional relativistic degrees of freedom from big-bang nucleosynthesis (BBN)
introduce the important restrictions. However MDM can decouple before the QCD phase transition and evade these bounds~\cite{Sigurdson:2004zp}.
In any case, it is difficult to find a production mechanism associated to this
light MDM in order to account for the total amount of DM. In addition, there are
more constraining bounds for very light MDM even if it just constitutes part of the total non-baryonic matter. For example, the energy-loss analysis of stellar objects in globular clusters constraints dipole moments more strongly for masses $m_{\rm DM}\lesssim5$~keV \cite{raffelt}. Similar bounds can be found by taking into account the data from the supernova 1987A \cite{Kadota:2014mea}, but in this case, it can be extended up to masses of order $m_{\rm DM}\lesssim 10$~MeV or even $100$~MeV depending on different assumptions about the thermal properties of the
supernova.

For the {\it middle} region, with $10$~MeV $\lesssim m_{\rm DM} \lesssim 1$~GeV, the experimental and observational
constraints may be satisfied for a larger value of ${\mu_{\rm DM}}$.
In this case, the most robust constraints come from precision measurements and, in
particular, from the contribution of the MDM
to the running of the fine-structure constant, which modifies the mass of
the $W^\pm$ boson predicted in the SM~\cite{Sigurdson:2004zp}.
Similar constraints can be placed by MDM direct production in particle accelerators \cite{colliders}.
The cleaner environment makes the single photon channel data at LEP slightly more constraining than mono-jet signatures at the Tevatron or at the LHC \cite{Fortin:2011hv}.

For values close to the above bound, MDM can achieve the abundance by
the classical thermal freeze-out mechanism in order to account for the total DM density \cite{Sigurdson:2004zp, Fortin:2011hv}.
However, this type of DM suffers the constraints associated to general light WIMPs, and it
is difficult to think that DM with masses below $10$~GeV can constitute the total
missing matter (due to restrictions coming from observations of cosmic-ray positrons,
cosmic-ray antiprotons and radio observations \cite{Bringmann:2014lpa}).

Finally, for MDM heavier than $\sim 1$ GeV, the constraints on the value
of ${\mu_{\rm DM}}$ are even more important due to direct detection experiments. However,
in this and the former case, the DM abundance can be produced by the thermal freeze-in mechanism.
Within the standard inflationary framework, the preferred value for $\mu_{\rm DM}$ depends on
the reheat and maximum temperatures with respect to the MDM mass \cite{Aranda:2014kea}.

By the definition of MDM, the magnetic interaction with photons is its leading interaction
with SM particles. However, the possible values for the magnetic moment commented above
can be different if we assume a more involved cosmological setup, for example the relic abundance
can be larger if exotic processes increase the expansion rate during freeze-out \cite{KamTur90},
or if there is a particle-antiparticle asymmetry for the MDM.
Other laboratory constraints, as the one coming from the
Lamb shift \cite{millichargeone,davidson} or the targeted experiment at SLAC
\cite{slac}, are also subdominant. Astrophysical analyses related to the
stability of the Galactic disk, annihilations
in the solar neighborhood, or lifetimes of compact objects, are not competitive either
 \cite{Starkman:1990nj, Sigurdson:2004zp, cosmics}. The same situation was found by \cite{Sigurdson:2004zp} for the constraints derived from
Large-Scale Structure or the Cosmic Microwave Background. In contrast, different conclusions
can be found for the indirect signatures of MDM as we have already commented,
although there are important uncertainties involved in these studies
associated with different assumptions \cite{Jungman:1995df, Sigurdson:2004zp, Cembranos:2012nj}.

One motivated way of having MDM arises for composite DM~\cite{DiazCruz:2007be,Antipin:2014qva,Antipin:2015xia}. If that is the case, it is possible that the
constituents that form these {\it dark hadrons} might have non zero electric charge and thus contribute to
a non zero dipole magnetic moment for the bound or composite state.
Motivated by the situation in QCD, where the use of an $SU(3)_F$ flavor symmetry at low energies facilitates a description of the different mesons and baryons, we consider a similar situation for DM, where new fundamental fermionic degrees of freedom are introduced and assumed to interact strongly through an unspecified new  interaction present at a high energy scale. At lower energies, a flavor symmetry is assumed to exist that allows us to consider the different composite states to be analyzed in terms of their possible values for $\mu_{\rm DM}$. The new fundamental fermionic particles, denoted by $q$ in analogy to quarks, can be electrically charged and thus contain $SU(2)_L \times U(1)_Y$ quantum numbers.

Following the example of QCD, we think of the new strong interaction as a scaled-up version of it and thus consider, at low energies, a situation where a $SU(3)_D$ {\it dark} flavor symmetry is present for three new dark quarks that transform in its fundamental representation {\bf 3}. With respect to the Standard Model (SM) gauge group, they are singlets of $SU(3)_C$ and might transform non-trivially under $SU(2)_L \times U(1)_Y$. We then have the following possibilities: they can form a triplet of $SU(2)_L$ with one hypercharge $Y=y_1$; two of them can form a doublet and the third one a singlet with hypercharges $Y=y_1$, and $Y=y_2$, respectively; or they can all be singlets with independent hypercharges. We analyze each case separately.

The paper is organized as follows: in section~\ref{sec:composite} we present how the three new fermions lead to composite states associated to the dark flavor symmetry $SU(3)_D$. After these are presented, in Section~\ref{sec:neutral-states} we explore the different possibilities for the $SU(2)_L \times U(1)_Y$ quantum numbers  of the new states in order to determine if there are neutral composite states that can play the role of DM. Section~\ref{sec:magnetic-moments}  shows the expressions for ${\mu_{\rm DM}}$ in all cases considered. In Section~\ref{sec:gell-mann-nishijima}, we discuss a generalized version of the Gell-Mann-Nishijima relation for our case, and finally, we conclude in Section~\ref{sec:conclusions}.

\section{Composite states from $SU(3)_D$}
\label{sec:composite}
\begin{figure}[H]
 \centering
 \includegraphics[scale=0.5]{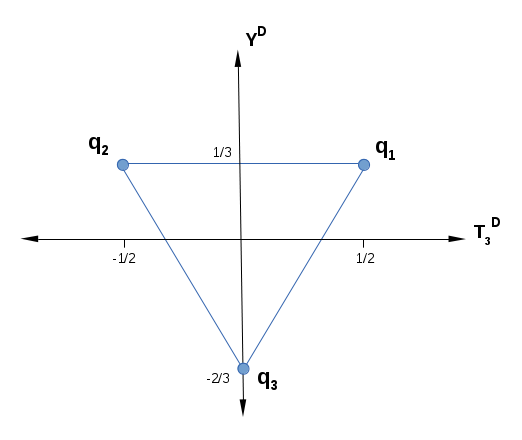}
 \caption{$SU(3)_D$ triplet of elementary particles.}
\label{fig:qi-YDvsT3D}
\end{figure}

First, we describe how three fundamental fermions in the {\bf 3} of $SU(3)_D$ can form the composite states. Using the fact that they belong to the fundamental representation, the new elementary particles can be characterized by their $T_3^D$ and $Y_D$ "quantum numbers", corresponding to the eigenvalues of the two diagonal generators of $SU(3)_D$ (see Appendix~\ref{appendix:QCD} for the QCD case). If we denote each fundamental state by $q_i=q(Y_D, T_3^D)$,
then we have (see Figure~\ref{fig:qi-YDvsT3D})

\begin{equation}
 q_1 = q\left(\tfrac{1}{3}, \tfrac{1}{2}\right), \hspace*{1cm} q_2 = q\left(\tfrac{1}{3}, -\tfrac{1}{2}\right),
 \hspace*{1cm} q_3 = q\left(-\tfrac{2}{3}, 0\right).
\end{equation}

Composite states made up of three constituents are obtained in the triple product of the fundamental representation ${\bf3}$ (see Appendix~\ref{appendix:wavefunctions} for the description of the spin wavefunctions used in the analysis),

\begin{equation}
 {\bf3}\otimes{\bf3}\otimes{\bf3}={\bf1}\oplus{\bf8}\oplus{\bf8}\oplus{\bf10} .
\end{equation}

\begin{figure}[H]
 \centering
 \includegraphics[scale=0.6]{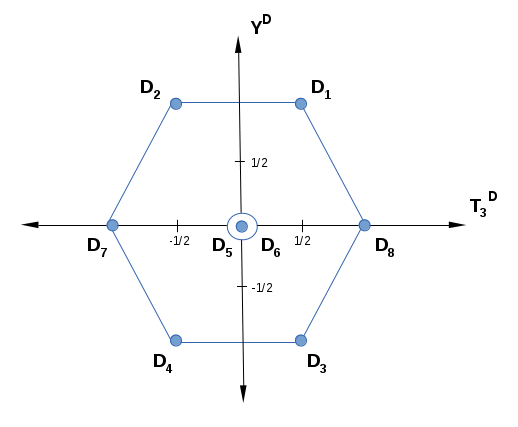}
 \caption{Octet states.}
\label{fig:octet}
\end{figure}

The octet states are denoted by $D_i(q_k, q_l, q_m)=D(Y^D, T_3^D)$, with $i=1,2...,8$ and $k,l,m=1,2,3$. From this notation, we have (see Figure~\ref{fig:octet})
\begin{eqnarray}\nonumber
\label{octet-states}
 D_1(q_1, q_1, q_2) & = & D\left(1, \tfrac{1}{2}\right), \quad
 D_2(q_1, q_2, q_2) = D\left(1, -\tfrac{1}{2}\right), \quad
 D_3(q_1, q_3, q_3) = D\left(-1, \tfrac{1}{2}\right), \quad \\ \nonumber
 D_4(q_2, q_3, q_3) &=& D\left(-1, -\tfrac{1}{2}\right), \quad
 D_5(q_1, q_2, q_3) = D\left(0, 0\right), \quad
 D_6(q_1, q_2, q_3) = D\left(0, 0\right), \\
 D_7(q_2, q_2, q_3) &=& D\left(0, -1\right), \quad
 D_8(q_1, q_1, q_3) = D\left(0, 1\right).
\end{eqnarray}

\begin{figure}[H]
 \centering
 \includegraphics[scale=0.6]{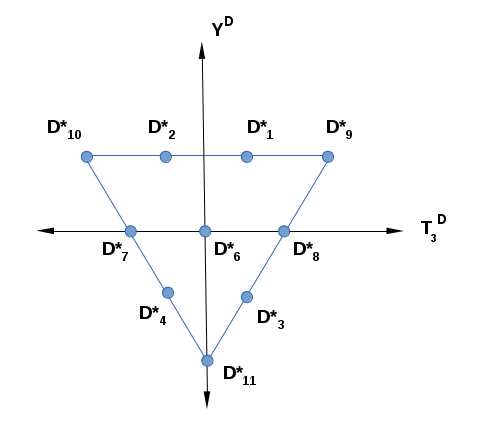}
 \caption{Decuplet states.}
\label{fig:decuplet}
\end{figure}

In the same way we can denote the decuplet states  shown in Figure~\ref{fig:decuplet} as $D^*_i(q_k, q_l, q_m)=D^*(Y^D, T_3^D)$:
\begin{eqnarray} \nonumber
\begin{tabular}{lll}
 $D_1^*(q_1, q_1, q_2) = D^*\left(1, \tfrac{1}{2}\right)$,&
 $D_2^*(q_1, q_2, q_2) = D^*\left(1, -\tfrac{1}{2}\right)$, &
 $D_3^*(q_1, q_3, q_3) = D^*\left(-1, \tfrac{1}{2}\right)$, \\
 $D_4^*(q_2, q_3, q_3) = D^*\left(-1, -\tfrac{1}{2}\right)$, &
 $D_6^*(q_1, q_2, q_3) = D^*\left(0, 0\right)$, &
 $D_7^*(q_2, q_2, q_3) = D^*\left(0, -1\right)$,\\
 $D_8^*(q_1, q_1, q_3) = D^*\left(0, 1\right)$, &
 $D_9^*(q_1, q_1, q_1) = D^*\left(1, \tfrac{3}{2}\right)$, &
 $D_{10}^*(q_2, q_2, q_2) = D^*\left(1, -\tfrac{3}{2}\right)$, \\
 $D_{11}^*(q_3, q_3, q_3) = D^*\left(-2, 0\right)$.
\end{tabular}
\\
\end{eqnarray}

These are the composite states that we consider in this work. The next step is to explore the different possibilities emanating from the different $SU(2)_L \times U(1)_Y$ assignments of the new fermions $q_i$ in order to determine the neutral composite states. Note that we label one of the states of the decuplet with $D_{11}^*$ instead of $D_{5}^*$. In this way, we can relate the particles $D_{i}$  of the octet with the $D_{i}^*$  associated with the decuplet since they are formed by the same constituents, as we will discuss.

\section{Charge assignments and neutral states}
\label{sec:neutral-states}
Since we are interested in the electrically neutral composite states, we want to know the conditions under which these
states will be neutral for each of the three different charge assignments mentioned above, i.e. whether the three new fundamental fermions form a triplet, a doublet plus a singlet or three singlets of $SU(2)_L$. The electric charge of the fermion $f_i$ is then determined using the relation $Q_i= T_3 + Y/2$, where $T_3$ and $Y$ correspond to the third component of isospin and hypercharge of the fermion $f_i$, respectively.

Note that $D_i$ and $D^*_i$ are made up of the same particles for $i=1, \dots, 8$ and so, when specifying the neutral states, we only do it for $D_i$, $D^*_9$, $D^*_{10}$ and $D^*_{11}$.

It is important to note that at this level of discussion, we are assuming that there is a flavor symmetry $SU(3)_D$ that is present in the new sector at low energies. Although in principle, and as a first and natural extension, we do have in mind a scaled-up version of QCD, we do not explicitly consider the strongly interacting gauge group that is assumed to be present at high energy. 

\subsection{Triplet}
Let the three new fermions $q_i$ form a triplet of $SU(2)_L$ with $Y=y_1$. A priori, we have the freedom of assigning any $q_i$ in the $SU(2)_L$ triplet to any {\it position} in the $SU(3)_D$ triplet, however, as we will see later, the different combinations can be related and so, we use the simplest one in which they occupy the same position.

The electric charges of the fundamental particles $q_i$ are then given by
\begin{equation}\label{Q3}
 Q_1 = 1+\frac{y_1}{2}, \hspace*{1cm} Q_2 = \frac{y_1}{2}, \hspace*{1cm} Q_3 = -1+\frac{y_1}{2}.
\end{equation}

Using these charges  we find that the composite states will be neutral in the following situations:
\begin{enumerate}
 \item $D_1$ is neutral for $y_1 = -\frac{4}{3}$.
 \item $D_4$ is neutral for $y_1 = \frac{4}{3}$.
 \item $D_2$ and $D_8$ are neutral for $y_1 = -\frac{2}{3}$.
 \item $D_3$ and $D_7$ are neutral for $y_1 = \frac{2}{3}$.
 \item $D_5$, $D_6$ and $D_{10}^*$ are neutral for $y_1 = 0$.
 \item $D_9^*$ is neutral for $y_1 = -2$.
 \item $D_{11}^*$ is neutral. for $y_1 = 2$.
\end{enumerate}

Note that $D_9^*$, $D_{10}^*$, and $D_{11}^*$ are made up of three neutral fundamental particles and therefore they do not have magnetic moment (in a model where only  the valence contribution is considered).

\subsection{Doublet and singlet}
Consider now the case where we have a $SU(2)_L$ doublet and a singlet. Let the hypercharge of the doublet be $y_1$, and the one of the singlet be $y_2$. The corresponding electric charges are given by
\begin{equation}\label{Q21}
 Q_1 = \frac{1}{2}+\frac{y_1}{2}, \hspace*{1cm} Q_2 = -\frac{1}{2}+\frac{y_1}{2}, \hspace*{1cm} Q_3 = \frac{y_2}{2}.
\end{equation}

The composite states will be neutral in the following situations:
\begin{enumerate}
 \item $D_1$ is neutral if $y_1 = -\frac{1}{3}$, independently of $y_2$.
 \item $D_2$ is neutral if $y_1 = \frac{1}{3}$, independently of $y_2$.
 \item $D_3$ is neutral if $y_1 = -\left(1+2y_2\right)$.
 \item $D_4$ is neutral if $y_1 = 1-2y_2$.
 \item $D_5$ and $D_6$ are neutral if $y_1 = -\frac{y_2}{2}$.
 \item $D_7$ is neutral if $y_1 = 1-\frac{y_2}{2}$.
 \item $D_8$ is neutral if $y_1 = -\left(1+\frac{y_2}{2}\right)$.
 \item $D^*_9$ is neutral if $y_1 = -1$, independently of $y_2$.
 \item $D^*_{10}$ is neutral if $y_1 = 1$, independently of $y_2$.
 \item $D^*_{11}$ is neutral if $y_2 = 0$, independently of $y_1$.
\end{enumerate}

\subsection{Singlets}
Let the hypercharges of the three $SU(2)_L$ singlets be $y_1$, $y_2$, $y_3$, respectively. Their electric charges are given by
\begin{equation}\label{Q111}
 Q_1 = \frac{y_1}{2}, \hspace*{1cm} Q_2 = \frac{y_2}{2}, \hspace*{1cm} Q_3 = \frac{y_3}{2}.
\end{equation}

Now we find that the composites states will be neutral if at least one of the following conditions is satisfied:
\begin{enumerate}
 \item $D_1$ is neutral if $y_1 = -\frac{y_2}{2}$, independently of $y_3$.
 \item $D_2$ is neutral if $y_1 = -2y_2$, independently of $y_3$.
 \item $D_3$ is neutral if $y_1 = -2y_3$, independently of $y_2$.
 \item $D_4$ is neutral if $y_3 = -\frac{y_2}{2}$, independently of $y_1$.
 \item $D_5$ and $D_6$ are neutral if $y_1+y_2+y_3=0$.
 \item $D_7$ is neutral if $y_2 = -\frac{y_3}{2}$, independently of $y_1$.
 \item $D_8$ is neutral if $y_1 = -\frac{y_3}{2}$, independently of $y_2$.
 \item $D^*_9$, $D^*_{10}$ or $D^*_{11}$ are neutral if $y_1=0$, $y_2=0$ or $y_3=0$, respectively. However we are not
 interested in these conditions, since the states are made up of three neutral particles and therefore their
 magnetic moments is zero.
\end{enumerate}

Each one of the conditions in $1$-$4$, $7$ and $8$ can be considered individually or combined with one of the others
($1$ with $4$, $2$ with $3$, and $7$ with $8$) so that we get $y_i=y_j=-\frac{y_k}{2}\neq0$, for $i\neq j\neq k$. Then,
depending on the values of $y_1$, $y_2$ and $y_3$ it is possible to obtain two, three or four neutral states.

\section{Magnetic moments}
\label{sec:magnetic-moments}
The expressions for the magnetic moments of the octect and decuplet states, in terms of the magnetic moments of their constituents, are the same as those obtained in the Quark Model using the spin-flavor wavefunctions given in appendix~\ref{appendix:wavefunctions}. For the octet states the results are
\begin{equation}
\label{magnetic-octet}
\begin{tabular}{ll}
 $\mu_{D_1} = \tfrac{1}{3}\left(4\mu_1-\mu_2\right)$, & $\mu_{D_2} = \tfrac{1}{3}\left(4\mu_2-\mu_1\right)$, \\
 $\mu_{D_3} = \tfrac{1}{3}\left(4\mu_3-\mu_1\right)$, & $\mu_{D_4} = \tfrac{1}{3}\left(4\mu_3-\mu_2\right)$, \\
 $\mu_{D_5} = \mu_3$, &  $\mu_{D_6} = \tfrac{2}{3}\left(\mu_2+\mu_1\right)-\tfrac{1}{3}\mu_3$, \\
 $\mu_{D_7} = \tfrac{1}{3}\left(4\mu_2-\mu_3\right)$,& $\mu_{D_8} = \tfrac{1}{3}\left(4\mu_1-\mu_3\right)$ ,
\end{tabular}
\end{equation}
and for the decuplet states we get
\begin{equation}
\label{magnetic-decuplet}
\begin{tabular}{lll}
 $\mu_{D_1^*} = 2\mu_1+\mu_2$ , & $\mu_{D_2^*} = 2\mu_2+\mu_1$ , & $\mu_{D_3^*} =  2\mu_3+\mu_1$ , \\
 $\mu_{D_4^*} = 2\mu_3+\mu_2$ , & $\mu_{D_6^*} = \mu_1+\mu_2+\mu_3 $, & $\mu_{D_7^*} = 2\mu_2+\mu_3$ , \\
 $\mu_{D_8^*} = 2\mu_1+\mu_3$ , & $\mu_{D_9^*} = 3\mu_1$, & $\mu_{D_{10}^*} = 3\mu_2$ , \\ & $\mu_{D_{11}^*} = 3\mu_3$ . &
\end{tabular}
\end{equation}

We see that, except for $D_5$ and $D_6$, there are symmetries under the exchanges $i\leftrightarrow j$, $j\leftrightarrow k$, $k\leftrightarrow i$, and $i\to j \to k \to i$: the same magnetic moments are obtained among different states. Consider for example the case $1 \leftrightarrow 2$ and note that $\mu_{D_1} \leftrightarrow \mu_{D_2}$. The origin of these relations can be traced to the symmetries of the spin-wavefunctions and this, in fact, is the reason of why we obtain the same magnetic moments (though corresponding to different symmetry-related states), independently of how we assign the order between the components in the $SU(2)_L$ and $SU(3)_D$ multiplets.

Defining the mass ratios $r_{1j}\equiv m_1/m_j$, we can express all the magnetic moments in units of $\frac{e\hbar}{2m_1}$. We obtain the results shown in Tables~\ref{table:triplet} ,\ref{table:doublet-singlet} and \ref{table:singlets} for the magnetic moments of the neutral dark hadrons when the constituents are in a $SU(2)_L$ triplet, doublet plus singlet and three singlets respectively. The first column presents the hypercharges leading to the neutral composite states shown in the second column. The third column displays the expressions for the magnetic moments in units of $\frac{e\hbar}{2m_1}$. The last three columns are added for curiosity: they show specific values for degeneracies in the constituents masses. Note that the zero magnetic moments in these three last columns are {\it accidental} since they exist only in the case of exact mass degeneracy.

Recall that there is an ordering symmetry among the components in the multiplets of $SU(2)_L$ and $SU(3)_D$ except  for $D_5$ and $D_6$. In this case the way we relate the components of the multiplets matters. We show this in Tables~\ref{table:tildetriplet}, \ref{table:tildedoublet}, and \ref{table:tildesinglets}, where we use the following notation: call
$\tilde q_i$ the components of the $SU(2)_L$ multiplet and $q_i$ those of the $SU(3)_D$ one (our previous analysis and results in Tables~\ref{table:triplet} ,\ref{table:doublet-singlet} and \ref{table:singlets} corresponds to the case $\tilde{q_i} = q_i$ for the $SU(2)_L$). The first column of each of these tables contains the relation among the multiplets, followed by the expressions for the magnetic moments.


\begin{table}[H]
\centering
{\renewcommand{\arraystretch}{1.25}
\renewcommand{\tabcolsep}{0.20cm}
\begin{tabular}{||p{3.25cm}|p{1.2cm}|p{2.5cm}|c|c|c||}
 \hline
 Hypercharge ($y_i$) Charge ($Q_1$,$Q_2$,$Q_3$)  &  Dark hadron & $\mu_{D_i,D^*_i}$  $\left(\frac{e\hbar}{2m_1}\right)$ & $r_{12}=1$ & $r_{13}=1$ & $r_{12}=r_{13}$ \\
 \hline \hline
 $y_1=-4/3$ & $D_1$   & $\frac{2}{9}\left(2+r_{12}\right)$ & $2/3$ & $-$ &$-$ \\
     $(1/3, -2/3,-5/3)$ & $D_1^*$ & $\frac{2}{3}\left(1-r_{12}\right)$ & $0^{**}$ &$-$ & $-$\\
 \hline
 $y_1=4/3$   & $D_4$   & $-\frac{2}{9}\left(2r_{13}+r_{12}\right)$ & $-$ & $-$ & $-\frac{2}{3}r_{12}$ \\
$(5/3, 2/3, -1/3)$    & $D_4^*$ & $\frac{2}{3} \left(r_{12}-r_{13}\right)$ &$-$ &$-$ & $0^{**}$ \\
 \hline
 & $D_2$   & $-\frac{2}{9}\left(1+2r_{12}\right)$ & $-2/3$  &$-$ & $-$ \\
 $y_1=-2/3$  & $D_8$   & $\frac{4}{9} \left(2+r_{13}\right)$ &$-$ & $4/3$ &$-$ \\
$(2/3, -1/3,-4/3)$           & $D_2^*$ & $\frac{2}{3}\left(1-r_{12}\right)$ & $0^{**}$ & $-$&$-$ \\
& $D_8^*$ & $\frac{4}{3}\left(1-r_{13}\right)$ & $-$ & $0^{**}$ & $-$\\
 \hline
& $D_3$   & $-\frac{4}{9}\left(1+2r_{13}\right)$ & $-$ & $-4/3$ & $-$ \\
 $y_1=2/3$    & $D_7$   & $\frac{2}{9}\left(r_{13}+2r_{12}\right)$ &$-$ &$-$ &  $\frac{2}{3}r_{12}$ \\
$(4/3,1/3,-2/3)$ & $D_3^*$ & $\frac{4}{3}\left(1-r_{13}\right)$ & $-$ & $0^{**}$ & $-$ \\
& $D_7^*$ & $\frac{2}{3}\left(r_{12}-r_{13}\right)$ & $-$ & $-$ & $0^{**}$ \\
 \hline
& $D_5$   & $-r_{13}$ & $-$ & $-1$ & $-$ \\
$y_1=0$& $D_6$   & $\frac{1}{3}\left(2+r_{13}\right)$ & $-$ & $1$ & $-$ \\
$(1,0,-1)$& $D_6^*$ & $1-r_{13}$ & $-$ & $0^{**}$ & $-$ \\
& $D_{10}^*$ & $0^*$ & $-$ & $-$ & $-$ \\
 \hline
 $y_1=-2$   & $D_{9}^*$  & $0^*$ & $-$ & $-$ & $-$ \\
$(0,-1,-2)$    & & & & &\\
 \hline
 $y_1=2$& $D_{11}^*$ & $0^*$ & $-$ & $-$ & $-$ \\
$(2,1,0)$   & & & & & \\
 \hline
\end{tabular}}
\caption{Magnetic moments of the neutral states if the three constituents are in a $SU(2)_L$ triplet. The last three columns correspond to specific cases where there are mass degeneracies among the different constituents ($r_{1j}\equiv m_1/m_j$). Note that the cases where the magnetic moment is zero fall in two different categories: those that are zero because their constituents are neutral (denoted by $0^*$) and those where a degeneracy in constituents mass is present (denoted by $0^{**}$).}
\label{table:triplet}
\end{table}

\begin{table}[H]
\centering
{\renewcommand{\arraystretch}{1.25}
\renewcommand{\tabcolsep}{0.2cm}
\begin{tabular}{||p{3.25cm}|p{1.2cm}|p{3.2cm}|c|c|c||}
 \hline
 Hypercharge ($y_i$) Charge ($Q_1$,$Q_2$,$Q_3$)  &  Dark hadron & $\mu_{D_i,D^*_i}$  $\left(\frac{e\hbar}{2m_1}\right)$ & $r_{12}=1$ & $r_{13}=1$ & $r_{12}=r_{13}$ \\
 \hline \hline
$y_1=-1/3$, $y_2$ & $D_1$   & $\frac{2}{9} \left(2+r_{12}\right)$  & $2/3$  & $-$ & $-$ \\
$(1/3,-2/3,y_2/2)$ & $D_1^*$ & $\frac{2}{3}\left(1-r_{12}\right)$ & $0^{**}$  & $-$ & $-$ \\
\hline
 $y_1=1/3$, $y_2$ & $D_2$   & $-\frac{2}{9}\left(1+2r_{12}\right)$  & $-2/3$ & $-$ & $-$ \\
 $(2/3,-1/3,y_2/2)$ & $D_2^*$ & $\frac{2}{3}\left(1-r_{12}\right)$ & $0^{**}$  & $-$ & $-$ \\
\hline
 $y_1=-\left(1+2y_2\right)$ & $D_3$   & $\frac{y_2}{3}\left(1+2r_{13}\right)$ & $-$ & $y_2$ & $-$ \\
 $(-y_2,-(1+y_2),\frac{y_2}{2})$ & $D_3^*$ & $-y_2\left(1-r_{13}\right)$ & $-$ & $0^{**}$ & $-$\\
\hline
 $y_1=\left(1-2y_2\right)$& $D_4$   & $\frac{y_2}{3}\left(2r_{13}+r_{12}\right)$ & $-$ & $-$ & $y_2r_{13}$ \\
 $(1-y_2,-y_2,y_2/2)$ & $D_4^*$ & $y_2\left(r_{13}-r_{12}\right)$ & $-$ & $-$ & $0^{**}$\\
 \hline
 & $D_5$   & $\frac{y_2}{2}r_{13}$ & $-$ & $y_2/2$ & $-$ \\
 $y_1=-y_2/2$ & $D_6$   & $\frac{1}{3}\left(1-r_{12}\right)$   & & & \\
$(\frac{(2-y_2)}{4},\frac{-(2+y_2)}{4},\frac{y_2}{2})$ & & $-\frac{y_2}{6}\left(1+r_{12}+r_{13}\right)$ & $-$ & $-$ & $-$ \\
& $D_6^*$ & $\frac{1}{2}\left(1-r_{12}\right)$ & & & \\
& & $-\frac{y_2}{4}\left(1+r_{12}-2r_{13}\right)$ & $-$ & $-$ & $-$ \\
\hline
 $y_1=1-y_2/2$ & $D_7$   & $-\frac{y_2}{6}\left(r_{13}+2r_{12}\right)$  & $-$ & $-$ & $-y_2r_{13}/2$ \\
 $(1-\frac{y_2}{4}, \frac{-y_2}{4}, \frac{y_2}{2})$ & $D_7^*$ & $\frac{y_2}{2}\left(r_{13}-r_{12}\right)$ & $-$ & $-$ & $0^{**}$ \\
\hline
 $y_1=-\left(1+y_2/2 \right)$& $D_8$   & $-\frac{y_2}{6}\left(2+r_{13}\right)$ & $-$ & $-y_2/2$ & $-$ \\
 $(-\frac{y_2}{4},-(1+\frac{y_2}{4}),\frac{y_2}{2})$ & $D_8^*$ & $-\frac{y_2}{2}\left(1-r_{13}\right)$ & $-$ & $0^{**}$ & $-$ \\
   \hline
 $y_1=-1$, $y_2$& $D_9^*$ & $0^*$ & $-$ & $-$ & $-$ \\
 $(0,-1,y_2/2)$ & & & & & \\
\hline
 $y_1=1$, $y_2$& $D_{10}^*$ & $0^*$ & $-$ & $-$ & $-$ \\
 $(1,0,y_2/2)$ & & & & & \\
 \hline
 $y_1$, $y_2=0$ & $D_{11}^*$ & $0^*$ & $-$ & $-$ & $-$ \\
 $(\frac{\left(y_1+1\right)}{2},\frac{\left(y_1-1\right)}{2},0)$ & & & & & \\
 \hline
\hline
\end{tabular}}
\caption{Magnetic moments of the neutral states when the new fermions form a $SU(2)_L$ doublet and a singlet. The last three columns correspond to specific cases where there are mass degeneracies among the different constituents ($r_{1j}\equiv m_1/m_j$). Note that the cases where the magnetic moment is zero fall in two different categories: those that are zero because their constituents are neutral (denoted by $0^*$) and those where a degeneracy in constituents mass is present (denoted by $0^{**}$).}
\label{table:doublet-singlet}
\end{table}

\begin{table}[H]
\centering
{\renewcommand{\arraystretch}{1.25}
\renewcommand{\tabcolsep}{0.2cm}
\begin{tabular}{||p{3.25cm}|p{1.2cm}|p{3.2cm}|c|c|c||}
 \hline
 Hypercharge ($y_i$) Charge ($Q_1$,$Q_2$,$Q_3$)  &  Dark hadron & $\mu_{D_i,D^*_i}$  $\left(\frac{e\hbar}{2m_1}\right)$ & $r_{12}=1$ & $r_{13}=1$ & $r_{12}=r_{13}$ \\
 \hline \hline
 $y_1=-y_2/2$, $y_3$ & $D_1$   & $-\frac{y_2}{6}\left(2+r_{12}\right)$   & $-y_2/2$ & $-$ & $-$ \\
 $(-y_2/4,y_2/2,y_3/2)$ & $D_1^*$ & $-\frac{y_2}{2}\left(1-r_{12}\right)$ & $0^{**}$ & $-$ & $-$ \\
 \hline
 $y_1=-2y_2$, $y_3$ & $D_2$   & $\frac{y_2}{3}\left(1+2r_{12}\right)$ & $y_2$ & $-$ & $-$ \\
 $(-y_2,y_2/2,y_3/2)$ & $D_2^*$ & $-y_2\left(1-r_{12}\right)$& $0^{**}$   & $-$ & $-$ \\
 \hline
 $y_1=-2y_3$, $y_2$ & $D_3$   & $\frac{y_3}{3}\left(1+2r_{13}\right)$ & $-$ & $y_3$ & $-$ \\
 $(-y_3,y_2/2,y_3/2)$ & $D_3^*$ & $-y_3\left(1-r_{13}\right)$& $-$ & $0^{**}$   & $-$\\
 \hline
 $y_1$, $y_3=-y_2/2$& $D_4$   & $-\frac{y_2}{6}\left(r_{12}+2r_{13}\right)$  & $-$ & $-$ & $-y_2r_{13}/2$ \\
 $(y_1/2,y_2/2,-y_2/4)$ & $D_4^*$ & $\frac{y_2}{2}\left(r_{12}-r_{13}\right)$ & $-$ & $-$ & $0^{**}$ \\
 \hline
 & $D_5$   & $y_3r_{13}/2$ & $-$ & $y_3/2$ & $-$ \\
 $y_1+y_2+y_3=0$  & $D_6$   & $\frac{y_2\left(r_{12}-1\right)}{3}-\frac{y_3\left(2+r_{13}\right)}{6}$ & $-$ & $-$ & $-$\\
$(\frac{-(y_2+y_3)}{2},\frac{y_2}{2},\frac{y_3}{2})$ & $D_6^*$ & $\frac{y_2\left(r_{12}-1\right)}{2}-\frac{y_3\left(1-r_{13}\right)}{2}
$ & $-$ & $-$ & $-$ \\
 \hline
 $y_1$, $y_3=-2y_2$& $D_7$   & $\frac{y_2}{3}\left(2r_{12}+r_{13}\right)$  & $-$ & $-$ & $ y_2r_{13}$ \\
 $(y_1/2,y_2/2,-y_2)$ & $D_7^*$ & $y_2\left(r_{12}-r_{13}\right)$& $-$ & $-$ & $0^{**}$ \\
 \hline
 $y_1=-y_3/2$, $y_2$ & $D_8$   & $-\frac{y_3}{6}\left(2+r_{13}\right)$ & $-$ & $-y_3/2$ &  $-$ \\
 $(-y_3/4,y_2/2,y_3/2)$ & $D_8^*$ & $-\frac{y_3}{2}\left(1-r_{13}\right)$ & $-$ & $0^{**}$ & $-$ \\
 \hline
 $y_1=0$, $y_2$, $y_3$& $D_9^*$ & $0^*$ & $-$ & $-$ & $-$ \\
 $(0,y_2/2,y_3/2) $ & & & & & \\
 \hline
 $y_1$, $y_2=0$, $y_3$  & $D_{10}^*$ & $0^*$ & $-$ & $-$ & $-$ \\
 $(y_1/2,0,y_3/2)$ & & & & & \\
 \hline
 $y_1$, $y_2$, $y_3=0$ & $D_{11}^*$ & $0^*$ & $-$ & $-$ & $-$ \\
 $(y_1/2,y_2/2,0)$ & & & & & \\
 \hline
\end{tabular}}
\caption{Magnetic moments of the neutral states if each of the three constituents is a $SU(2)_L$ singlet. The last three columns correspond to specific cases where there are mass degeneracies among the different constituents ($r_{1j}\equiv m_1/m_j$). Note that the cases where the magnetic moment is zero fall in two different categories: those that are zero because their constituents are neutral (denoted by $0^*$) and those where a degeneracy in constituents mass is present (denoted by $0^{**}$).}
\label{table:singlets}
\end{table}


\begin{table}[H]
 \centering
{\renewcommand{\arraystretch}{1.25}
\renewcommand{\tabcolsep}{0.2cm}
\begin{tabular}{||p{5cm}|p{5.5cm}||}
 \hline
 Relation among multiplets & Magnetic moments $\mu_{D_{5,6}}$  $\left(\frac{e\hbar}{2m_1}\right)$  \\
 \hline
\hline
 $\tilde q_1=q_2$, $\tilde q_2=q_1$, $\tilde q_3=q_3$ & $\mu_{D_5}=-r_{13}$  \\
& $\mu_{D_6}=\frac{1}{3}\left(2r_{12}+r_{13}\right)$  \\
 \hline
 $\tilde q_1=q_3$, $\tilde q_2=q_2$, $\tilde q_3=q_1$ & $\mu_{D_5}=r_{13}$ \\
                                    & $\mu_{D_6}=-\frac{1}{3}\left(2+r_{13}\right)$  \\
 \hline
 $\tilde q_1=q_1$, $\tilde q_2=q_3$, $\tilde q_3=q_2$ & $\mu_{D_5}=0^*$ \\
                                    & $\mu_{D_6}=\frac{2}{3}\left(1-r_{12}\right)$  \\
 \hline
 $\tilde q_1=q_2$, $\tilde q_2=q_3$, $\tilde q_3=q_1$ & $\mu_{D_5}=0^*$ \\
                                                      & $\mu_{D_6}=-\frac{2}{3}\left(1-r_{12}\right)$  \\
 \hline
 $\tilde q_1=q_3$, $\tilde q_2=q_1$, $\tilde q_3=q_2$ & $\mu_{D_5}=r_{13}$ \\
                                                      & $\mu_{D_6}=-\frac{1}{3}\left(2r_{12}+r_{13}\right)$  \\
 \hline
\end{tabular}}
\caption{Magnetic moments of the states $D_5$ and $D_6$ if the three constituents are in a $SU(2)_L$ triplet for different relations among the multiplet components (see text for definition of $\tilde q_i$). The case $\tilde q_i =q_i$ is included in Table~\ref{table:triplet}. $r_{1j}\equiv m_1/m_j$.}
\label{table:tildetriplet}
\end{table}

\begin{table}[H]
\centering
{\renewcommand{\arraystretch}{1.25}
\renewcommand{\tabcolsep}{0.2cm}
\begin{tabular}{||p{5cm}|p{5.5cm}||}
 \hline
 Relation among multiplets & Magnetic moments $\mu_{D_{5,6}}$  $\left(\frac{e\hbar}{2m_1}\right)$  \\
 \hline
\hline
 $\tilde q_1=q_2$, $\tilde q_2=q_1$, $\tilde q_3=q_3$ & $\mu_{D_5}=y_2 r_{13}/2$ \\
& $\mu_{D_6}=-\frac{\left(1-r_{12}\right)}{3}-\frac{y_2\left(1+r_{12}+r_{13}\right)}{6}$ \\
 \hline
 $\tilde q_1=q_3$, $\tilde q_2=q_2$, $\tilde q_3=q_1$ & $\mu_{D_5}=\left(2-y_2\right)r_{13}/4$ \\
& $\mu_{D_6}=-\frac{\left(r_{13}+2r_{12}\right)}{6}+\frac{y_2\left(4+r_{13}-2r_{12}\right)}{12}$ \\
 \hline
 $\tilde q_1=q_1$, $\tilde q_2=q_3$, $\tilde q_3=q_2$ & $\mu_{D_5}=-\left(2+y_2\right)r_{13}/4$ \\
                                    & $\mu_{D_6}=\frac{\left(2+r_{13}\right)}{6}+\frac{y_2\left(r_{13}+4r_{12}-2\right)}{12}$ \\
 \hline
 $\tilde q_1=q_2$, $\tilde q_2=q_3$, $\tilde q_3=q_1$ & $\mu_{D_5}=-\left(2+y_2\right)r_{13}/4$ \\
& $\mu_{D_6}=\frac{\left(r_{13}+2r_{12}\right)}{6}+\frac{y_2\left(4+r_{13}-2r_{12}\right)}{12}$ \\
 \hline
 $\tilde q_1=q_3$, $\tilde q_2=q_1$, $\tilde q_3=q_2$ & $\mu_{D_5}=\left(2-y_2\right)r_{13}/4$ \\
& $\mu_{D_6}=\frac{-\left(2+r_{13}\right)}{6}+\frac{y_2\left(r_{13}+4r_{12}-2\right)}{12}$ \\
 \hline
\end{tabular}}
\caption{Magnetic moments of the states $D_5$ and $D_6$ if the three constituents are in a $SU(2)_L$ doublet plus singlet for different relations among the multiplet components (see text for definition of $\tilde q_i$). The case $\tilde q_i =q_i$ is included in Table~\ref{table:doublet-singlet}. $r_{1j}\equiv m_1/m_j$.}
\label{table:tildedoublet}
\end{table}

\begin{table}[H]
\centering
{\renewcommand{\arraystretch}{1.25}
\renewcommand{\tabcolsep}{0.2cm}
\begin{tabular}{||p{5cm}|p{5.5cm}||}
 \hline
 Relation among multiplets & Magnetic moments $\mu_{D_{5,6}}$  $\left(\frac{e\hbar}{2m_1}\right)$  \\
 \hline
\hline
 $\tilde q_1=q_2$, $\tilde q_2=q_1$, $\tilde q_3=q_3$ & $\mu_{D_5}=y_3r_{13}/2$ \\
& $\mu_{D_6}=\frac{y_2\left(1-r_{12}\right)}{3}-\frac{y_3\left(2r_{12}+r_{13}\right)}{6}$ \\
 \hline
 $\tilde q_1=q_3$, $\tilde q_2=q_2$, $\tilde q_3=q_1$ & $\mu_{D_5}=-\left(y_2+y_3\right)r_{13}/2$ \\
& $\mu_{D_6}=\frac{y_2\left(2r_{12}+r_{13}\right)}{6}+\frac{y_3\left(2+r_{13}\right)}{6}$ \\
 \hline
$\tilde q_1=q_1$, $\tilde q_2=q_3$, $\tilde q_3=q_2$ & $\mu_{D_5}=y_2r_{13}/2$ \\
& $\mu_{D_6}=-\frac{y_2\left(2+r_{13}\right)}{6}-\frac{y_3\left(1-r_{12}\right)}{3}$ \\
 \hline
 $\tilde q_1=q_2$, $\tilde q_2=q_3$, $\tilde q_3=q_1$ & $\mu_{D_5}=y_2r_{13}/2$ \\
                                                      & $\mu_{D_6}=-\frac{y_2\left(2r_{12}+r_{13}\right)}{6}+\frac{y_3\left(1-r_{12}\right)}{3}$ \\
 \hline
 $\tilde q_1=q_3$, $\tilde q_2=q_1$, $\tilde q_3=q_2$ & $\mu_{D_5}=-\left(y_2+y_3\right)r_{13}/2$ \\
                                                      & $\mu_{D_6}=\frac{y_2\left(2+r_{13}\right)}{6}+\frac{y_3\left(2r_{12}+r_{13}\right)}{6}$ \\
 \hline
\end{tabular}}
\caption{Magnetic moments of the states $D_5$ and $D_6$ if the three constituents are singlets of $SU(2)_L$ for different relations among the multiplet components (see text for definition of $\tilde q_i$). The case $\tilde q_i =q_i$ is included in Table~\ref{table:singlets}. $r_{1j}\equiv m_1/m_j$.}
\label{table:tildesinglets}
\end{table}

\section{Gell-Mann-Nishijima formula}
\label{sec:gell-mann-nishijima}
The electric charge of a particle is given by
\begin{equation}
 Q = T_3 + \frac{1}{2}Y_W,
\end{equation}
where $T_3$ and $Y_W$ are the third component of isospin and the hypercharge, associated to the generators of the gauge groups $SU(2)_L$
and $U(1)_Y$, respectively.

In the Quark Model, where it is considered a flavor $SU(3)_F$ symmetry, there are
two group diagonal generators, $I_3$ and $Y_F$. As it turns out, because the diagonal generators in this case correspond to $SU(2)$ and $U(1)$ symmetries, the specific assignment of the light quarks under the $SU(3)_F$ as a fundamental allows us to express electric charge in terms of the $SU(3)_F$ generators through the well known Gell-Mann-Nishijima formula
\begin{equation}
  Q = I_3 + \frac{1}{2}Y_F.
\end{equation}

Since in our model we consider three fundamental particles with arbitrary electric charge, it is of interest to determine those cases where it is possible to define the electric charge in terms of the flavor generators $T_3^D$, $Y^D$ of the group
$SU(3)_D$, as in the Quark Model, with a generalization of the Gell-Mann-Nishijima formula
$Q = c_TT_3^D + c_YY^D$. This of course depends crucially on the relation between $\tilde q_i$ and $q_i$ and the different $SU(3)_D$ representations used above.  We now discuss each case separately.

\subsection{Triplet}
When the generalization can be used, the charges are given by
\begin{equation}\label{Q_i}
 Q_i = \frac{c_T}{2} + \frac{c_Y}{3}, \hspace*{1cm} Q_j = -\frac{c_T}{2} + \frac{c_Y}{3}, \hspace*{1cm}
 Q_k = -\frac{2}{3}c_Y, \hspace*{1cm} i\neq j\neq k.
\end{equation}

Note that here there are only two independent relations while in (\ref{Q3}) the three relations are dependent. If we want to express the charge in terms of the flavor generators, the system of equations is only consistent for $y_1=0$. For this particular value of the hypercharge we obtain four neutral states, $D_5$,
$D_6$, $D_6^*$ and $D_{10}^*$.
\begin{itemize}
 \item For $\tilde q_i=q_i$, $c_Y=\frac{3}{2}$ and $c_T=1$.
 \item For $\tilde q_1=q_2$, $\tilde q_2=q_1$, and $\tilde q_3=q_3$, $c_Y=\frac{3}{2}$ and $c_T=-1$.
 \item For $\tilde q_1=q_3$, $\tilde q_2=q_2$, and $\tilde q_3=q_1$, $c_Y=-\frac{3}{2}$ and $c_T=-1$.
 \item For $\tilde q_1=q_1$, $\tilde q_2=q_3$, and $\tilde q_3=q_2$, $c_Y=0$ and $c_T=2$.
 \item For $\tilde q_1=q_2$, $\tilde q_2=q_3$, and $\tilde q_3=q_1$, $c_Y=0$ and $c_T=-2$.
 \item For $\tilde q_1=q_3$, $\tilde q_2=q_1$, and $\tilde q_3=q_2$, $c_Y=-\frac{3}{2}$ and $c_T=1$.
\end{itemize}

\subsection{Doublet and singlet}
From the charges given in equation (\ref{Q21}), the generalization can be used for two relations between $y_1$ and
$y_2$:
\begin{enumerate}
 \item $y_2=-2y_1$ can be used with the following relations between $\tilde q_i$ and $q_i$:
\begin{itemize}
 \item $\tilde q_i=q_i$, $c_Y=\frac{3}{2}y_1$ and $c_T=1$.
 \item $\tilde q_1=q_2$, $\tilde q_2=q_1$ and $\tilde q_3=q_3$, $c_Y=\frac{3}{2}y_1$ and $c_T=1$.
 \item $\tilde q_1=q_3$, $\tilde q_2=q_2$ and $\tilde q_3=q_1$, $c_Y=-\frac{3}{4}\left(y_1+1\right)$ and
       $c_T=-\frac{1}{2}\left(3y_1-1\right)$.
 \item $\tilde q_1=q_1$, $\tilde q_2=q_3$ and $\tilde q_3=q_2$, $c_Y=-\frac{3}{4}\left(y_1-1\right)$ and
       $c_T=\frac{1}{2}\left(3y_1+1\right)$.
 \item $\tilde q_1=q_3$, $\tilde q_2=q_1$ and $\tilde q_3=q_2$, $c_Y=-\frac{3}{4}\left(y_1+1\right)$ and
       $c_T=\frac{1}{2}\left(3y_1-1\right)$.
\end{itemize}
Note that  $y_2=-2y_1$ is consistent with almost all the necessary conditions to obtain neutral states, except for
 $D_7$ and $D_8$ when $\tilde q_i=q_i$.

\item $y_2=y_1+1$ works when $\tilde q_1=q_2$, $\tilde q_2=q_3$ and $\tilde q_3=q_1$,
$c_Y=-\frac{3}{4}\left(y_1-1\right)$ and \\ $c_T=-\frac{1}{2}\left(3y_1+1\right)$. This relation between the
hypercharges is consistent with all the conditions for neutral states.
\end{enumerate}

\subsection{Singlets}
Here we have that each particle $\tilde q_i$ has a charge given by $\frac{y_i}{2}$, so independently of the relation
between $\tilde q_i$ and $q_i$, we obtain the same relations between the three hypercharges, up to a change of the
form $i\rightarrow j$, $j\rightarrow k$, $k\rightarrow i$.\\

For the particular relation $\tilde q_i=q_i$, it is found that the generalization can be used when

\begin{equation}
 c_Y = \tfrac{3}{4}\left(y_1+y_2\right), \hspace*{1cm} c_T = \tfrac{1}{2}\left(y_1-y_2\right) = \tfrac{1}{2}\left(y_2-y_1\right),
\end{equation}
this is
\begin{equation}
 y_1 = y_2, \hspace*{1cm} y_3 = -2y_1, \hspace*{1cm} c_Y = \tfrac{3}{2}y_1, \hspace*{1cm} c_T = 0.
\end{equation}

These relations are satisfied by the conditions for the neutral states $D_5$, $D_6$, $D_7$ and $D_8$. The conditions
for the remaining states could also be satisfied, but this takes place only for $y_1=y_2=y_3=0$.

\section{Conclusions}
\label{sec:conclusions}
MDM posses an interesting and useful possibility that broadens up the spectrum of candidates and scenarios for the problem of DM. In this work, we consider DM candidates that can have a magnetic dipole moment due to the fact that they are composite states coming from a high energy strongly interacting sector.

Three additional elementary fermions have been introduced in addition to the SM particle content. These new fermions are singlets under $SU(3)_C$ but can have non trivial representations under the electroweak gauge group. Since there are three of them, then only three different possibilities for their $SU(2)_L$ transformation exist: triplet, doublet plus singlet or three singlets.

Assuming there is a low energy $SU(3)_D$ {\it dark flavor symmetry}, and in analogy with low energy QCD, we construct the composite states and determine those that can be neutral. We find that there are several possibilities for each of the three cases above. 

The results for $\mu_{\rm DM}$ are presented in terms of the constituents masses and hypercharges, and cases where they are zero are singled out for two different scenarios: i) when the constituents themselves are electrically neutral and ii) when there is a particular mass degeneracy among some of the elementary constituents masses.

\appendix
\section{QCD} \label{appendix:QCD}

QCD is a particular case for the charge assignment corresponding to one $SU(2)_L$ doublet and one singlet with
$\tilde q_i=q_i$, $y_1=\frac{1}{3}$, $y_2=-2y_1=-\frac{2}{3}$, $c_Y=\frac{3}{2}y_1=\frac{1}{2}$ and $c_T=1$. Here the
three fundamental particles $q_i$ correspond to the three light quarks

\begin{equation}
 q_1\rightarrow u, \hspace*{1cm} q_2\rightarrow d, \hspace*{1cm} q_3\rightarrow s,
\end{equation}

which carry a charge

\begin{equation}
 Q_u = \frac{1}{2}+\frac{y_1}{2} = \frac{2}{3}, \hspace*{1cm} Q_d = -\frac{1}{2}+\frac{y_1}{2} = -\frac{1}{3},
 \hspace*{1cm} Q_s = \frac{y_2}{2} = -\frac{1}{3}.
\end{equation}

For the octet states the correspondence is

\begin{equation}
\begin{split}
 &D_1\rightarrow p, \hspace*{1cm} D_2\rightarrow n, \hspace*{1cm} D_3\rightarrow \Xi^0, \hspace*{1cm}
 D_4\rightarrow \Xi^-, \\\\
 &D_5\rightarrow \Lambda, \hspace*{1cm} D_6\rightarrow \Sigma^0, \hspace*{1cm} D_7\rightarrow \Sigma^-,
 \hspace*{1cm} D_8\rightarrow \Sigma^+,
\end{split}
\end{equation}

while for the decuplet states we have

\begin{equation}
\begin{split}
 &D_1^*\rightarrow \Delta^{+}, \hspace*{1cm} D_2^*\rightarrow \Delta^{0}, \hspace*{1cm} D_3^*\rightarrow \Xi^{*0},
 \hspace*{1cm} D_4^*\rightarrow \Xi^{*-}, \hspace*{1cm} D_6^*\rightarrow \Sigma^{*0}, \\\\
 & D_7^*\rightarrow \Sigma^{*-}, \hspace*{1cm} D_8^*\rightarrow \Sigma^{*+}, \hspace*{1cm} D_9^*\rightarrow\Delta^{++},
 \hspace*{1cm} D_{10}^*\rightarrow\Delta^{-} \hspace*{1cm} D_{11}^* \rightarrow \Omega^-.
\end{split}
\end{equation}

\section{Spin-flavor wavefunctions}\label{appendix:wavefunctions}

If we consider the spin as an internal degree of freedom, then each fundamental particle $q_i$ is in the spin-flavor
group product $SU(3)\otimes SU(2)$. So the composite states are obtained in the product

\begin{equation}
 \left({\bf3}\otimes{\bf3}\otimes{\bf3}\right)\otimes\left({\bf2}\otimes{\bf2}\otimes{\bf2}\right).
\end{equation}

The decomposition of this product leads to singlets, octects and decuplets with spin $1/2$ and $3/2$.
The spin-flavor wavefunctions of the octet states are~\cite{Thirring:1965pka}

\begin{equation}
\begin{split}
 \Ket{D_{1\shortuparrow}} =& \tfrac{1}{\sqrt{18}}\Big(
 2\Ket{q_{1\shortuparrow} q_{1\shortuparrow} q_{2\shortdownarrow}} + 2\Ket{q_{1\shortuparrow} q_{2\shortdownarrow} q_{1\shortuparrow}} +
 2\Ket{q_{2\shortdownarrow} q_{1\shortuparrow} q_{1\shortuparrow}} - \Ket{q_{1\shortuparrow} q_{1\shortdownarrow} q_{2\shortuparrow}} -
 \Ket{q_{1\shortuparrow} q_{2\shortuparrow} q_{1\shortdownarrow}} - \Ket{q_{2\shortuparrow} q_{1\shortuparrow} q_{1\shortdownarrow}}
 \\&-\Ket{q_{1\shortdownarrow} q_{1\shortuparrow} q_{2\shortuparrow}} - \Ket{q_{1\shortdownarrow} q_{2\shortuparrow} q_{1\shortuparrow}} -
 \Ket{q_{2\shortuparrow} q_{1\shortdownarrow} q_{1\shortuparrow}} \Big)
 \\ =&\tfrac{1}{\sqrt{18}}\Big(2\Ket{q_{1\shortuparrow} q_{1\shortuparrow} q_{2\shortdownarrow}} -
 \Ket{q_{1\shortuparrow} q_{1\shortdownarrow} q_{2\shortuparrow}} -\Ket{q_{1\shortdownarrow} q_{1\shortuparrow} q_{2\shortuparrow}}
 + permutations\Big),
 \\\\
 \Ket{D_{2\shortuparrow}} =& \tfrac{1}{\sqrt{18}}\Big(2\Ket{q_{2\shortuparrow} q_{2\shortuparrow} q_{1\shortdownarrow}} -
 \Ket{q_{2\shortuparrow} q_{2\shortdownarrow} q_{1\shortuparrow}} - \Ket{q_{2\shortdownarrow} q_{2\shortuparrow} q_{1\shortuparrow}} +
 permutations \Big) ,
 \\\\
 \Ket{D_{3\shortuparrow}} =& \tfrac{1}{\sqrt{18}}\Big(2\Ket{q_{3\shortuparrow} q_{3\shortuparrow} q_{1\shortdownarrow}} -
 \Ket{q_{3\shortuparrow} q_{3\shortdownarrow} q_{1\shortuparrow}} - \Ket{q_{3\shortdownarrow} q_{3\shortuparrow} q_{1\shortuparrow}} +
 permutations \Big) ,
 \\\\
 \Ket{D_{4\shortuparrow}} =& \tfrac{1}{\sqrt{18}}\Big(2\Ket{q_{3\shortuparrow} q_{3\shortuparrow} q_{2\shortdownarrow}} -
 \Ket{q_{3\shortuparrow} q_{3\shortdownarrow} q_{2\shortuparrow}} - \Ket{q_{3\shortdownarrow} q_{3\shortuparrow} q_{2\shortuparrow}} +
 permutations \Big) ,
 \\\\
 \Ket{D_{5\shortuparrow}} =& \tfrac{1}{\sqrt{12}}\Big(\Ket{q_{2\shortuparrow} q_{3\shortuparrow} q_{1\shortdownarrow}} +
 \Ket{q_{3\shortuparrow} q_{2\shortuparrow} q_{1\shortdownarrow}} - \Ket{q_{3\shortuparrow} q_{1\shortuparrow} q_{2\shortdownarrow}} -
 \Ket{q_{1\shortuparrow} q_{3\shortuparrow} q_{2\shortdownarrow}} + permutations \Big) .
 \\\\
 \Ket{D_{6\shortuparrow}} =& \tfrac{1}{6}\Big(2\Ket{q_{2\shortuparrow} q_{1\shortuparrow} q_{3\shortdownarrow}} +
 2\Ket{q_{1\shortuparrow} q_{2\shortuparrow} q_{3\shortdownarrow}} - \Ket{q_{3\shortuparrow} q_{2\shortuparrow} q_{1\shortdownarrow}} -
 \Ket{q_{3\shortuparrow} q_{1\shortuparrow} q_{2\shortdownarrow}} - \Ket{q_{1\shortuparrow} q_{3\shortuparrow} q_{2\shortdownarrow}} -
 \Ket{q_{1\shortdownarrow} q_{3\shortuparrow} q_{2\shortuparrow}} + permutations \Big) ,
 \\\\
  \Ket{D_{7\shortuparrow}} =& \tfrac{1}{\sqrt{18}}\Big(2\Ket{q_{2\shortuparrow} q_{2\shortuparrow} q_{3\shortdownarrow}} -
 \Ket{q_{2\shortuparrow} q_{2\shortdownarrow} q_{3\shortuparrow}}-\Ket{q_{2\shortdownarrow} q_{2\shortuparrow} q_{3\shortuparrow}} +
 permutations \Big) ,
 \\\\
 \Ket{D_{8\shortuparrow}} =& \tfrac{1}{\sqrt{18}}\Big(2\Ket{q_{1\shortuparrow} q_{1\shortuparrow} q_{3\shortdownarrow}} -
 \Ket{q_{1\shortuparrow} q_{1\shortdownarrow} q_{3\shortuparrow}}-\Ket{q_{1\shortdownarrow} q_{1\shortuparrow} q_{3\shortuparrow}} +
 permutations \Big) .
\end{split}
\end{equation}

In a similar way, for the decuplet states the wavefunctions are given by

\begin{equation}
\begin{split}
 D_{1\shortuparrow}^* =& \tfrac{1}{\sqrt{3}}\Big(\Ket{q_{1\shortuparrow} q_{1\shortuparrow} q_{2\shortuparrow}} +
 permutations\Big) ,
 \hspace*{1cm}
 D_{2\shortuparrow}^* = \tfrac{1}{\sqrt{3}}\Big(\Ket{q_{2\shortuparrow} q_{2\shortuparrow} q_{1\shortuparrow}} +
 permutations\Big) ,
 \\\\
 D_{3\shortuparrow}^* =& \tfrac{1}{\sqrt{3}}\Big(\Ket{q_{3\shortuparrow} q_{3\shortuparrow} q_{2\shortuparrow}} +
 permutations\Big) ,
 \hspace*{1cm}
 D_{4\shortuparrow}^* = \tfrac{1}{\sqrt{3}}\Big(\Ket{q_{3\shortuparrow} q_{3\shortuparrow} q_{2\shortuparrow}} +
 permutations\Big) ,
 \\\\
 D_{6\shortuparrow}^* =& \tfrac{1}{\sqrt{6}}\Big(\Ket{q_{3\shortuparrow} q_{2\shortuparrow} q_{1\shortuparrow}} +
 \Ket{q_{2\shortuparrow} q_{3\shortuparrow} q_{1\shortuparrow}} + permutations\Big) ,
 \\\\
 D_{7\shortuparrow}^* =& \tfrac{1}{\sqrt{3}}\Big(\Ket{q_{2\shortuparrow} q_{2\shortuparrow} q_{3\shortuparrow}} +
 permutations\Big) ,
 \hspace*{1cm}
 D_{8\shortuparrow}^* = \tfrac{1}{\sqrt{3}}\Big(\Ket{q_{1\shortuparrow} q_{1\shortuparrow} q_{3\shortuparrow}} +
 permutations\Big) ,
 \\\\
 D_{9\shortuparrow}^* =& \Ket{q_{1\shortuparrow} q_{1\shortuparrow} q_{1\shortuparrow}} ,
 \hspace*{2.5cm}
 D_{10\shortuparrow}^* = \Ket{q_{2\shortuparrow} q_{2\shortuparrow} q_{2\shortuparrow}} ,
 \hspace*{2.5cm}
 D_{11\shortuparrow}^* = \Ket{q_{3\shortuparrow} q_{3\shortuparrow} q_{3\shortuparrow}} ,
\end{split}
\end{equation}
where $permutations$ indicates the change $1\rightarrow3$ and $2\rightarrow3$ over each of the previous kets.

\begin{acknowledgments}
This work has been supported in part by SNI and CONACYT CB-2011-01-167425 (Fondo Sectorial de Investigaci\'on para la Educaci\'on - M\'exico).
J.A.R.C. acknowledges financial support from the MINECO (Spain) projects FIS2014-52837-P, FPA2014-53375-C2-1-P,
{\it Programa Becas Iberoam\'erica} funded by Santander Universidades (Spain) 2015, and Consolider-Ingenio MULTIDARK CSD2009-00064.
\end{acknowledgments}




\end{document}